\documentclass[a4paper,10pt]{article}
\usepackage{amssymb,amsfonts,amsmath,mathtext,cite,enumerate,float,mathrsfs}
\usepackage{graphicx,color}
\usepackage{wrapfig}
\graphicspath{{pic/}}

\usepackage{epsfig}
\usepackage{epsf}
\usepackage{subfig} 
\usepackage{feynmp} 
\usepackage{pgf,pgfarrows,pgfnodes,pgfautomata,pgfheaps,multirow}
\usepackage{array}
\usepackage{bm}
\usepackage{latexsym}
\usepackage{pifont}

\title{$\tau^{-} \to \pi^{-} \pi^{0} \nu_\tau$ decay in the extended NJL model}

\author{M. K. Volkov\footnote{E-mail address: volkov@theor.jinr.ru}, D. G. Kostunin\footnote{E-mail address: kostunin@theor.jinr.ru}\\
\it Bogoliubov Laboratory of Theoretical Physics, JINR\\ 
\it Dubna, 141980, Russia}

\begin{document}

\maketitle

\begin{abstract}
The width of the decay $\tau^{-} \to \pi^{-} \pi^{0} \nu_\tau$ was 
calculated in the extended NJL model.
Contact interaction of $W$ boson with a pion pair as well as the contribution of 
the $\rho$ mesons in the ground and first radial-exited states are taken into account. 
The sum of the contact diagram and diagram with intermediate $\rho$ meson in the ground state 
leads to the result which coincides with the result of the vector-dominance model. 
Our results are in satisfactory agreement with experimental data.
\\

\\


12.39.Fe 	Chiral Lagrangians 
\end{abstract}



\section{Introduction}
The decay $\tau^{-} \to \pi^{-} \pi^{0} \nu_\tau$ is well-studied from experimental~\cite{PDG,cleo,aleph,belle} and theoretical~\cite{okun,kuhn,outlook,Jegerlehner:2011ti} points of view. A set of phenomenological models was used for a theoretical description of $\tau$ lepton decays. The chiral NJL model was also used by one of the authors of this paper~\cite{VolkovIvanovOsipov3pi,VolkovIvanovOsipovpigamma} particularly for a description of decays $\tau \to 3\pi \nu_\tau$ and $\tau \to \gamma \pi \nu_\tau$. In these works intermediate axial-vector $a_1$ and vector $\rho$ mesons in the ground state were taken into account. Recently, it was appeared the paper where the contribution of the intermediate $\rho'(1450)$ meson in an exited state was also took~\cite{tau3pi}. It made better agreement of theoretical results with experimental data in the interval of a pion pair invariant mass between 1.5 and 2 GeV~\cite{aleph}. 

In recent experiments, the influence of the intermediate radial-exited $\rho$ mesons on the decay $\tau^{-} \to \pi^{-} \pi^{0} \nu_\tau$ was included in  the description of experimental data~\cite{cleo,aleph,belle}. The Kuhn-Santamaria model~\cite{kuhn} was used for treatment of experimental data.

In this paper, a theoretical description of the decay $\tau^{-} \to \pi^{-} \pi^{0} \nu_\tau$ was given in the framework of the NJL model with intermediate vector mesons in the ground and first radial-excitation state. Firstly, the diagram with intermediate $W$ boson (contact diagram) and the diagram with the transition of $W$ boson into $\rho(770)$ meson in the standard NJL model were calculated~\cite{VolkovEbert,VolkovAn,pepan86,EbertReinhardt,pepan93,VolkovEbertReinhardt,UFN,HatsudaKunihiro,Vogl:1991qt}. 
Then an additional diagram with radial-exited $\rho'$ meson was calculated in the extended NJL model~\cite{VolkovWeiss,yadPh,VolkovEbertNagy,UFN}.  
Let us note that the result of calculation of the first two diagrams coincides with the results obtained in the vector-dominance model~\cite{outlook}. The contribution of the radial-exited intermediate $\rho$ meson to the amplitude of the decay $\tau \to \pi^{-} \pi^{0} \nu_\tau$ is also in satisfactory agreement with recent experimental data~\cite{cleo,aleph,belle}.

\section{Amplitude and width of $\tau \to \pi^{-} \pi^{0} \nu_\tau$ decay}

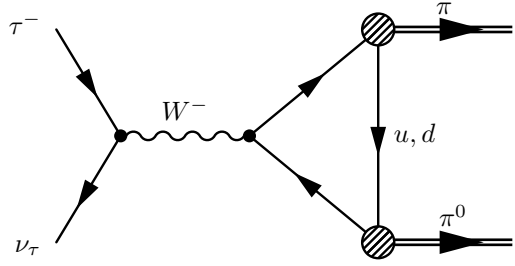
\begin{figure}[h]
\begin{center}

\begin{fmffile}{contact}

      \begin{fmfgraph*}(200,100)
	      \fmfpen{thin}\fmfleftn{l}{2}\fmfrightn{r}{2}

	      \fmfright{b,a}
	      \fmfleft{f,fb}
	      \fmflabel{$\tau^-$}{fb}
	      \fmflabel{$\nu_\tau$}{f}
	      \fmf{fermion}{fb,v1,f}
 	      \fmf{fermion,tension=.5}{p1,v2,p2}
	      \fmf{boson,lab.side=left,label=$W^{-}$}{v1,v2}
 	      \fmf{fermion,lab.side=left,label=$u,,d$}{p2,p1}	      
 	      \fmfdotn{v}{2}
	      \fmfblob{0.06w}{p2}
	      \fmfblob{0.06w}{p1}

	      \fmf{dbl_plain_arrow,lab.side=left,label=$\pi^{-}$}{p2,a}
 	      \fmf{dbl_plain_arrow,lab.side=left,label=$\pi^{0}$}{p1,b}

	      \fmfforce{120,90}{p2}
	      \fmfforce{120,10}{p1}

	      \fmfforce{170,90}{a}
	      \fmfforce{170,10}{b}

	      \fmfforce{0,90}{fb}
	      \fmfforce{0,10}{f}

      \end{fmfgraph*}

\end{fmffile}
\caption{Contact interaction $W^{-}$ boson with a pion pair}
\label{fig1}
\end{center}
\end{figure}

The amplitude of the $\tau \to \pi^{-} \pi^{0} \nu_\tau$ decay is described in the NJL model by given the Feynman diagrams in Figs.~\ref{fig1} and \ref{fig2}.

\begin{figure}[h]
\begin{center}

\begin{fmffile}{rho}

      \begin{fmfgraph*}(200,100)
	      \fmfpen{thin}\fmfleftn{l}{2}\fmfrightn{r}{2}

	      \fmfright{b,a}
	      \fmfleft{f,fb}
	      \fmflabel{$\tau^-$}{fb}
	      \fmflabel{$\nu_\tau$}{f}
	      \fmf{dbl_plain_arrow,lab.side=left,label=$\rho,,\rho^{'}$}{v1,v4}
	      \fmf{boson,label=$W^{-}$}{v2,v3}
	      \fmf{fermion,left,tension=.5}{v1,v2}
	      \fmf{fermion,label=$u,,d$,left,tension=.5}{v2,v1}
	      \fmf{fermion}{fb,v3,f}
 	      \fmf{fermion,tension=0.5}{p1,v4,p2}
 	      \fmf{fermion,lab.side=left,label=$u,,d$}{p2,p1}	      
	      \fmfdotn{v}{4}
	      \fmfblob{0.06w}{p2}
	      \fmfblob{0.06w}{p1}

	      \fmf{dbl_plain_arrow,lab.side=left,label=$\pi^{-}$}{p2,a}
 	      \fmf{dbl_plain_arrow,lab.side=left,label=$\pi^{0}$}{p1,b}

	      \fmfforce{220,90}{p2}
	      \fmfforce{220,10}{p1}

	      \fmfforce{270,90}{a}
	      \fmfforce{270,10}{b}

	      \fmfforce{0,90}{fb}
	      \fmfforce{0,10}{f}

      \end{fmfgraph*}

\end{fmffile}
\caption{Interaction with intermediate $\rho$ ($\rho'$) meson}
\label{fig2}
\end{center}
\end{figure}
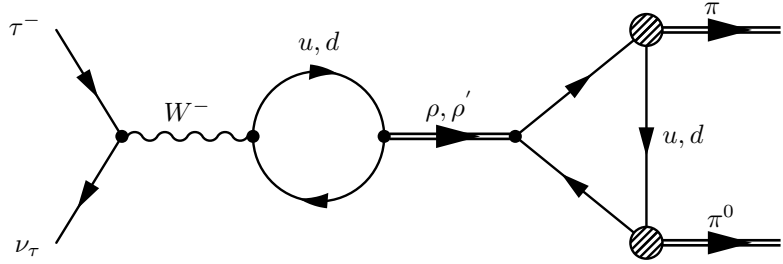

 \begin{figure}[h]
 \begin{center}
 \begin{tabular}{ccccccc}

	\begin{fmffile}{pi0}
      \begin{fmfgraph*}(63,48)

	      \fmfpen{thin}\fmfleftn{l}{2}\fmfrightn{r}{2}
 
	      \fmfright{b,a}
	      \fmfleft{f,fb}
 	      \fmf{vanilla}{fb,v1}
	      \fmf{vanilla,tension=.5}{p1,v1,p2}
	      \fmf{vanilla}{p2,p1}
	      \fmfblob{0.06w}{p2}
	      \fmfblob{0.06w}{p1}
	      \fmf{vanilla,lab.side=left,label= \begin{small} $\pi$ \end{small}}{p2,a}
	      \fmf{vanilla,lab.side=down,label=\begin{small}$\pi$\end{small}}{p1,b}

	      \fmfforce{38,43}{p2}
	      \fmfforce{38,3}{p1}

	      \fmfforce{63,43}{a}
	      \fmfforce{63,3}{b}

	      \fmfforce{0,23}{fb}

      \end{fmfgraph*}  
	\end{fmffile}

&  \multirow{1}[4]{*}[5ex]{$ = $ } 

&	\begin{fmffile}{pi1}
      \begin{fmfgraph*}(63,48)

	      \fmfpen{thin}\fmfleftn{l}{2}\fmfrightn{r}{2}
 
	      \fmfright{b,a}
	      \fmfleft{f,fb}
 	      \fmf{vanilla}{fb,v1}
	      \fmf{vanilla,tension=.5}{p1,v1,p2}
			\fmf{vanilla}{p2,p1}
	      \fmf{vanilla}{p2,a}
	      \fmf{vanilla}{p1,b}
   
	      \fmfforce{38,43}{p2}
	      \fmfforce{38,3}{p1}

	      \fmfforce{63,43}{a}
	      \fmfforce{63,3}{b}

	      \fmfforce{0,23}{fb}

      \end{fmfgraph*}
	\end{fmffile}

&  \multirow{1}[4]{*}[5ex]{$ + $ }


&	\begin{fmffile}{pi2}
      \begin{fmfgraph*}(63,48)

	      \fmfpen{thin}\fmfleftn{l}{2}\fmfrightn{r}{2}
 
	      \fmfright{b,a}
	      \fmfleft{f,fb}
 	      \fmf{vanilla}{fb,v1}
	      \fmf{vanilla,tension=.5}{p1,v1,p2}
	      \fmf{vanilla}{p2,a1}
	      \fmf{vanilla}{p1,a2}
	      \fmf{vanilla}{p1,p2}
	      \fmf{vanilla}{a2,b}
	      \fmf{vanilla}{a1,a}
	      \fmf{vanilla}{p2,a1}
	      \fmf{vanilla}{a1,a}
	      \fmf{vanilla,lab.side=down,label=$a_{1}$}{p1,a2}
	      \fmf{vanilla}{a2,b}
	      \fmf{dashes}{a2,a1}	%
	      \fmfv{decor.shape=circle,decor.filled=empty,decor.size=0.06w}{a2}
	      \fmfv{decor.shape=circle,decor.filled=30,decor.size=0.06w}{a1}
   
	      \fmfforce{48,43}{p2}
	      \fmfforce{48,3}{p1}
	      \fmfforce{60,43}{a1}	%
	      \fmfforce{60,3}{a2}	%
	      \fmfforce{73,43}{a}
	      \fmfforce{73,3}{b}

	      \fmfforce{0,23}{fb}

      \end{fmfgraph*}
	\end{fmffile}

&  \multirow{1}[4]{*}[5ex]{$ + $ }

&	\begin{fmffile}{pi3}
      \begin{fmfgraph*}(63,48)

	      \fmfpen{thin}\fmfleftn{l}{2}\fmfrightn{r}{2}
 
	      \fmfright{b,a}
	      \fmfleft{f,fb}
 	      \fmf{vanilla}{fb,v1}
	      \fmf{vanilla,tension=.5}{p1,v1,p2}
	      \fmf{vanilla}{p2,a1}
	      \fmf{vanilla}{p1,a2}
	      \fmf{vanilla}{p1,p2}
	      \fmf{vanilla}{a2,b}
	      \fmf{vanilla}{a1,a}
	      \fmf{vanilla,lab.side=left,label=$a_{1}$}{p2,a1}
	      \fmf{vanilla}{a1,a}
	      \fmf{vanilla,lab.side=down,label=$a_{1}$}{p1,a2}
	      \fmf{vanilla}{a2,b}

	      \fmfv{decor.shape=circle,decor.filled=empty,decor.size=0.06w}{a2}
	      \fmfv{decor.shape=circle,decor.filled=empty,decor.size=0.06w}{a1}
   
	      \fmfforce{38,43}{p2}
	      \fmfforce{38,3}{p1}

	      \fmfforce{63,43}{a}
	      \fmfforce{63,3}{b}

	      \fmfforce{0,23}{fb}

      \end{fmfgraph*}
	\end{fmffile}

 \end{tabular}
 \end{center}
\caption{Triangle diagrams with $\pi$ -- $a_1$ transitions}
 \end{figure}
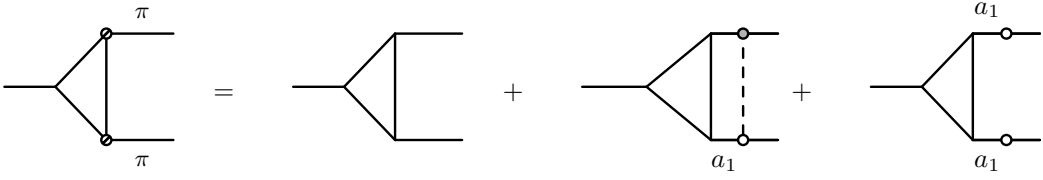

For description of the $W^{-} \pi^{-} \pi^{0}$ vertex one can use the result for the $\rho \to \pi \pi$ decay amplitude with the accounting of $\pi$ -- $a_1$ transitions. 
Amplitude of the $\rho \to \pi\pi$ is
\begin{equation}
g_\rho\left(Z + (1 - Z) + (f_{a_1}(p^2) - 1)\right)\rho^{-}_\mu (p_{\pi^{-}}^\mu - p_{\pi^{0}}^\mu) \pi^{-}\pi^{0}\, ,
\end{equation}
\begin{equation}
f_{a_1}(p^2) = 1 + \left(\frac{p^2 - m_\pi^2}{(g_\rho F_\pi)^2}\right)\left(1 - \frac{1}{Z}\right)\, ,
\end{equation}
where $p$ is the $\rho$ meson momentum, $p_{\pi^{-}}$ and $p_{\pi^{0}}$ are the outgoing pion momenta, $g_\rho \approx 6.14$ is the decay constant of $\rho \to \pi \pi$, $F_\pi = 93$ MeV is the pion decay constant, $Z = (1 - 6m_u^2/m_{a_1}^2)^{-1}$ is the additional renormalizing factor appeared after accounting of $\pi$ -- $a_1$ transitions, $m_u = 280$~MeV is the constituent quark mass, $m_{a_1} = 1230$~MeV is the mass of the $a_{1}$ meson, $m_\pi$ is the mass of the $\pi$ meson.

The first term of this amplitude is corresponding to triangle diagram without $\pi$ -- $a_1$ transitions, the second term is corresponding to diagram with $\pi$ -- $a_1$ transition on the one of the pion lines and the third term is corresponding to the diagram with transitions on the both pions lines\footnote{Let us note that the last term wasn't took into account in~\cite{pepan86}, because only constant terms were considered in that work.}.
For description of the $W^{-} \pi^{-} \pi^{0}$ vertex $g_\rho$  should be changed into $G_F |V_{ud}|$, where $G_F = 1.16637 \cdot 10^{-11} \: \mathrm{MeV^{-2}}$ is the Fermi constant; $|V_{ud}| = 0.97428$ is the Cabibbo angle cosine. For the first diagram we get
\begin{equation}
T_1 = G_F|V_{ud}| f_{a_1}(p^2) l_{\mu} (p_{\pi^{-}}^\mu - p_{\pi^{0}}^\mu) \pi^{-} \pi^{0}\, ,
\end{equation}
where $l_{\mu} = \bar{\nu}_\tau\gamma_\mu (1-\gamma^5)\tau$ is the lepton current.

The second diagram with the intermediate $\rho^{-}$ meson contains three parts.

The first part describes the $W^{-}$ into $\rho^{-}$ transition. For this part, one can use form describing a transition photon into $\rho$ meson calculated in~\cite{pepan86}. In this form, charge $e$ should be changed into $G_F|V_{ud}|$. We get
\begin{equation}
\frac{G_F|V_{ud}|}{g_\rho}(g^{\mu\nu}p^2-p^{\mu}p^{\nu})\, ,
\end{equation}
where $p = p_\tau - p_\nu$ is the $\rho$ meson momentum.

The $\rho$ meson propagator has the form
\begin{equation}
	\frac{g^{\mu\nu}}{m_{\rho}^2 - p^2 - i\sqrt{p^2}\Gamma_{\rho}(p^2)}\, ,
\end{equation}
where $\Gamma(m_\rho^2) = 149.1\: \mathrm{MeV}$ is the full width of the $\rho^{-}$ meson decay.

The last vertex corresponds to the $\rho^{-} \to \pi^{-}\pi^{0}$ decay through quark loop:
\begin{equation}
	g_\rho f_{a_1}(p^2)\rho^{-}_\mu (p_{\pi^{-}}^\mu - p_{\pi^{0}}^\mu)\pi^{-}\pi^{0}\, .
\end{equation}

This amplitude takes the form
\begin{equation}
T_2 = \frac{G_F|V_{ud}|f_{a_1}(p^2)p^2}{m_{\rho}^2 - p^2 - i\sqrt{p^2}\Gamma_{\rho}(p^2)} l_\mu (p_{\pi^{-}}^\mu - p_{\pi^{0}}^\mu) \pi^{-} \pi^{0}\, .
\end{equation}

The sum of these two diagrams has the form close to the vector meson dominance expression:
\begin{equation}
T_\rho = \frac{G_F|V_{ud}|f_{a_1}(p^2)m_\rho^2}{m_{\rho}^2 - p^2 - i\sqrt{p^2}\Gamma_{\rho}(p^2)}\left(1-i\frac{\sqrt{p^2}\Gamma_\rho(p^2)}{m_\rho^2}\right) l_\mu (p_{\pi^{-}}^\mu - p_{\pi^{0}}^\mu) \pi^{-} \pi^{0}\, .
\end{equation}

Let us consider the last part of the amplitude. It contains the intermediate radial-exited $\rho^{-}$ meson. The extended NJL model~\cite{VolkovWeiss, yadPh} should be used in this case. The probability of the transition of $W^{-}$ into $\rho^{-}(1450)$ meson can be calculated using the result for the photon transition into $\rho$ meson given in~\cite{ArbuzovKuraevVolkov}. Also, in this form charge $e$ should be changed into $G_F|V_{ud}|$. After that we should get:

\begin{equation}
	C_{W \rho'} \frac{G_F|V_{ud}|}{g_\rho}(g^{\mu\nu}p^2-p^{\mu}p^{\nu})\, ,
\end{equation}
\begin{equation}
	C_{W \rho'} = -\left(\frac{\cos(\beta+\beta_0)}{\sin(2\beta_0)} + \Gamma\frac{\cos(\beta - \beta_0)}{\sin{(2\beta_0)}}\right)\, ,
\end{equation}
where $\beta_0 = 61.44^{\circ}$, $\beta = 79.85^{\circ}$ are the mixing angles; $\Gamma = 0.54$ was defined in~\cite{VolkovEbertNagy}.

The propagator of the radial-exited $\rho$ meson reads:
\begin{equation}
	\frac{g^{\mu\nu}}{m_{\rho'}^2 - p^2 - i\sqrt{p^2}\Gamma_{\rho'}(p^2)}\, ,
\end{equation}
where $m_{\rho'} = 1465$ MeV is the mass of the radial-exited $\rho$ meson; $\Gamma_{\rho'}(m_{\rho'}^2) = 400$ MeV is its full width.

The $\rho' \to \pi\pi$ decay was considered in detail in~\cite{VolkovEbertNagy}. The amplitude of this process can be written:
\begin{equation}
	C_{\rho' \pi \pi}\rho^{-}_\mu f_{a_1}(p^2) (p_{\pi^{-}}^\mu - p_{\pi^{0}}^\mu) \pi^{-}\pi^{0}\, ,
\end{equation}
\begin{equation}
	C_{\rho' \pi \pi} = -\left(\frac{\cos(\beta + \beta_0)}{\sin(2\beta_0)}g_{\rho_1} + \frac{\cos(\beta - \beta_0)}{\sin(2\beta_0)}\frac{I^f_2}{I_2}g_{\rho_2}\right),
\end{equation}
where $g_{\rho_1} = g_{\rho} = 6.14$, $g_{\rho_2} = 10.56$ and definitions of $I^f_2$, $I_2$ were given in~\cite{VolkovEbertNagy}.

Our model can not describe relative phase between $\rho(770)$ and $\rho(1450)$. Thus, we should get phase from $e^{+}e^{-}$ annihilation and $\tau$ decays experiments: $T_{\rho'} \to e^{i\pi}T_{\rho'}$

For the part of the amplitude of the $\tau^{-} \to \pi^{-}\pi^{0} \nu_\tau$ decay containing the intermediate radial-exited $\rho$ meson one can get:
\begin{equation}
T_{\rho'} = e^{i\pi}\frac{G_F|V_{ud}|C_{W \rho'} C_{\rho' \pi \pi}(1/g_\rho) f_{a_1}(p^2) p^2}{m_{\rho'}^2 - p^2 - i\sqrt{p^2}\Gamma_{\rho'}(p^2)} (p_{\pi^{-}}^\mu - p_{\pi^{0}}^\mu) l_\mu \pi^{-} \pi^{0}\, .
\end{equation}

The sum of these three amplitudes is:
\begin{equation}
T = G_F|V_{ud}| f_{a_1}(p^2) m_\rho^2 \left( \frac{1-i\sqrt{q^2}\Gamma_\rho(p^2) / m_\rho^2}{m_{\rho}^2 - p^2 - i\sqrt{p^2}\Gamma_{\rho}(p^2)} + \frac{e^{i\pi}C_{W \rho'} C_{\rho' \pi \pi} (1/g_\rho) p^2/m_\rho^2}{m_{\rho'}^2 - p^2 - i\sqrt{p^2}\Gamma_{\rho'}(p^2)} \right) (p_{\pi^{-}}^\mu - p_{\pi^{0}}^\mu) l_\mu \pi^{-} \pi^{0}\, .
\end{equation}

After using the expression for the decay width we get $\mathcal{B}(\tau^{-} \to \pi^{-} \pi^{0} \nu_\tau) = 24.76$\%. This value is in satisfactory agreement with experimental data (see the Table~\ref{tbl1}).

The Kuhn-Santamaria model~\cite{kuhn} was used for treatment of experimental data. In this model the pion form-factor reads:
\begin{equation}
\frac{1}{1+\beta}\left(\frac{m_{\rho}^2}{m_{\rho}^2 - p^2 - i\sqrt{p^2}\Gamma_{\rho}(p^2)} + \beta \frac{ m_{\rho'}^2}{m_{\rho'}^2 - p^2 - i\sqrt{p^2}\Gamma_{\rho'}(p^2)} \right)\, ,
\end{equation}
where $\beta$ is a parameter taken from fitting of experimental data.

In our model we can get approximate value for $\beta$ at $p^2 = m_{\rho'}^2$
\begin{equation}
\beta \approx e^{i\pi}C_{W \rho'} C_{\rho' \pi \pi}/g_\rho \approx -0.086\,.
\end{equation}

This value is in satisfactory agreement with experimental data given in~\cite{cleo,aleph,belle} (see the Table~\ref{tbl1}).

We note that the $\rho'$ meson influence on the decay width is small. In fact, the contribution to the decay width from two first diagrams $\mathcal{B}_{1,2}(\tau^{-} \to \pi^{-} \pi^{0} \nu_\tau) = 24.68$\%. This value increased by $0.32$\% after the inclusion of the $\rho'$ meson contribution. This allowed us to describe differential width in the interval of the pion pair invariant mass above 1~GeV.

\begin{table}[h]
\begin{center}
\caption{Experimental and theoretical data}
{\begin{tabular}{@{}lllll@{}} 
\hline
 & CLEO~\cite{cleo} & ALEPH~\cite{aleph} & BELLE~\cite{belle} & Theory \\
\hline
 $\mathcal{B}(\tau^{-} \to \pi^{-} \pi^{0} \nu_\tau)$, \% & $25.32 \pm 0.15$ & $25.471 \pm 0.097 \pm 0.085$ & $25.24 \pm 0.01 \pm 0.39$ & $24.76 $ \\
 $\beta$ & $-0.108 \pm 0.007$ & $-0.097 \pm 0.006$  & $-0.15 \pm 0.05^{+0.15}_{-0.04}$  & $-0.086$\\ 
\hline
\end{tabular}}
\label{tbl1}
\end{center}
\end{table}

\section{Conclusions}

We have shown that results obtained in the framework of the NJL model satisfactorily describe the $\tau^{-} \to \pi^{-} \pi^{0} \nu_\tau$ decay.
This statement relates both the description of the partial decay width as well as the differential decay width. The value of the $\beta$ parameter computed in the NJL model framework is in quality agreement with one obtained from the fit of experimental data. It noticing that results were obtained in the NJL model with minimal number of parameters. We are going to describe processes $e^{+}e^{-} \to \pi\pi(\pi')$ and $\tau^{-} \to \pi^{-} \omega(\phi) \nu_\tau$ in the framework of the same model in future.

\section*{Acknowledgments}
We are grateful to E.~A.~Kuraev and A.~B.~Arbuzov for useful discussions. This work was supported by RFBR grant 10-02-01295-a.

\end{document}